\documentclass{svproc}
\usepackage[utf8]{inputenc}

\usepackage{url}

\usepackage[sectionbib]{natbib}
\bibpunct{(}{)}{;}{a}{}{,}%
\usepackage{graphicx}
\usepackage{amsmath,amssymb}
\usepackage{multirow}
\usepackage{longtable}

\usepackage[b5paper]{geometry}
\begin{document}
	\mainmatter              
	\title{Development of Large-Format Camera Systems \\ Based on the Latest Generation Sensors for the \\ 6-m Telescope}
	\titlerunning{Development of Large-Format Camera Systems}  
	%
	\author{V.I.~Ardilanov \and V.A.~Murzin \and I.V.~Afanasieva \and \\ N.G.~Ivaschenko \and M.A.~Pritychenko}
	\authorrunning{Ardilanov et al.} 
	%
	\tocauthor{Ardilanov~V.I., Murzin~V.A., Afanasieva~I.V., Ivaschenko~N.G., and Pritychenko~M.A.}
	\institute{Special Astrophysical Observatory, Russian Academy of Sciences, \\ Nizhnij Arkhyz, Russia,\\
		\email{valery@sao.ru},\\ WWW home page:
		\texttt{https://www.sao.ru/hq/adlab/}
		}
	\maketitle

\begin{abstract}
	The design and implementation of astronomical cameras based on the large-format CCD and CMOS detectors is described in this paper. The Dinacon-5 controller is used for work with the CCDs and to achieve high performance and low noise. A new controller is designed for CMOS sensors. The main characteristics of the provided systems are estimated on the basis of experimental data. The spatial autocorrelation analysis is applied for PSF estimation. The obtained test results are presented.
\end{abstract}

\section{Systems overview}
Astronomical camera systems (\cite{Murzin2016}) for the 6-meter telescope BTA have been built at the SAO RAS Advanced Design Laboratory over the past few years. The systems are based on the new types of large-format detectors:

\begin{itemize}
\item[--] deep-depleted $4K\times4K$ CCD231-84 and fully-depleted $2K\times4K$ CCD261-84 sensors with high response in the red and near-infrared spectral ranges;
\item[--] high-speed CMOS detectors GSense4040CMT and GSense6060BSI.
\end{itemize}

The new liquid nitrogen cryostats with enlarged optical windows and improved performance characteristics are implemented for the CCD systems.

A CCD-based system is handled by the universal low-noise Dinacon-5 controller, which is used for operating both single detectors and mosaic ones. 

A distinctive feature of the controller is the ability to drive CCD with thick substrate, requiring high back-substrate bias voltage. 

The CCD controller has a low-noise switching power supply module that is compact and lightweight. The communication module transfers the data to a computer using fiber-optic 1 Gigabit Ethernet interface.

The CCD systems main characteristics are presented in Table 1.

\begin{table}
\vspace{-2mm}
\caption{Specification of the CCD camera systems}
\begin{center}
	\begin{tabular}{l@{\quad}rl@{\quad}rl}
		\hline
		\multicolumn{1}{l}{\rule{0pt}{12pt}Parameter}&\multicolumn{2}{l}{CCD261-84}&\multicolumn{2}{l}{CCD231-84} \\[2pt]
		\hline\rule{0pt}{12pt}
		\hspace{-1mm}Image size, $pixels$  & $2080\times4112$ & & $4128\times4112$ & \\
		Pixel size, $\mu$m  & $15\times15$ & & $15\times15$ & \\
		Quantum efficiency (QE), $\%$  & $94 @800nm$  & & $94 @750nm$  & \\
		Readout rate, $Kpix/s$  & 65/185/335 & & 135/380/700 & \\
		Readout noise, $e^-$  & 2.18 & & 2.48 & \\
		Dynamic range  & 62800 & & 75000 & \\
		Nonlinearity, $\%$  & $<$0.1 & & $<$0.1 & \\
		Dark current, $e^-/pix/s$  & 0.001 & & 0.0005 & \\
		[2pt]
		\hline
	\end{tabular}
\end{center}
\end{table}

\vspace{-2mm}

For the system with a 200$\mu$m thick CCD261-84 we have investigated the magnitude of the fringing of incident and reflected waves in the near-infrared range in comparison with thinner detectors (\cite{Afanasiev2005, Afanasiev2011}). The fringing value for this CCD is 1.7\%,  which is more than 10 times less than for the CCD42-40 with a standard substrate thickness of 16$\mu$m.

Also, for this system, a point spread function (PSF) is investigated, which is expressed in the spreading of charge between neighboring pixels during exposure. The methodology proposed in \cite{Downing2006} is taken as the basis, and is expanded by introducing additional measurements of the degree of spreading for the red and blue ranges of light and the back-substrate bias voltage of the CCD261-84. The results of this study are shown in Figure 1 and reflect the presence of the effect of charge spreading at the different control voltages, being the basis for choosing the optimal voltage on the back substrate of the detector.

\begin{figure}
	\vspace{-3mm}
	\includegraphics[width=\columnwidth]{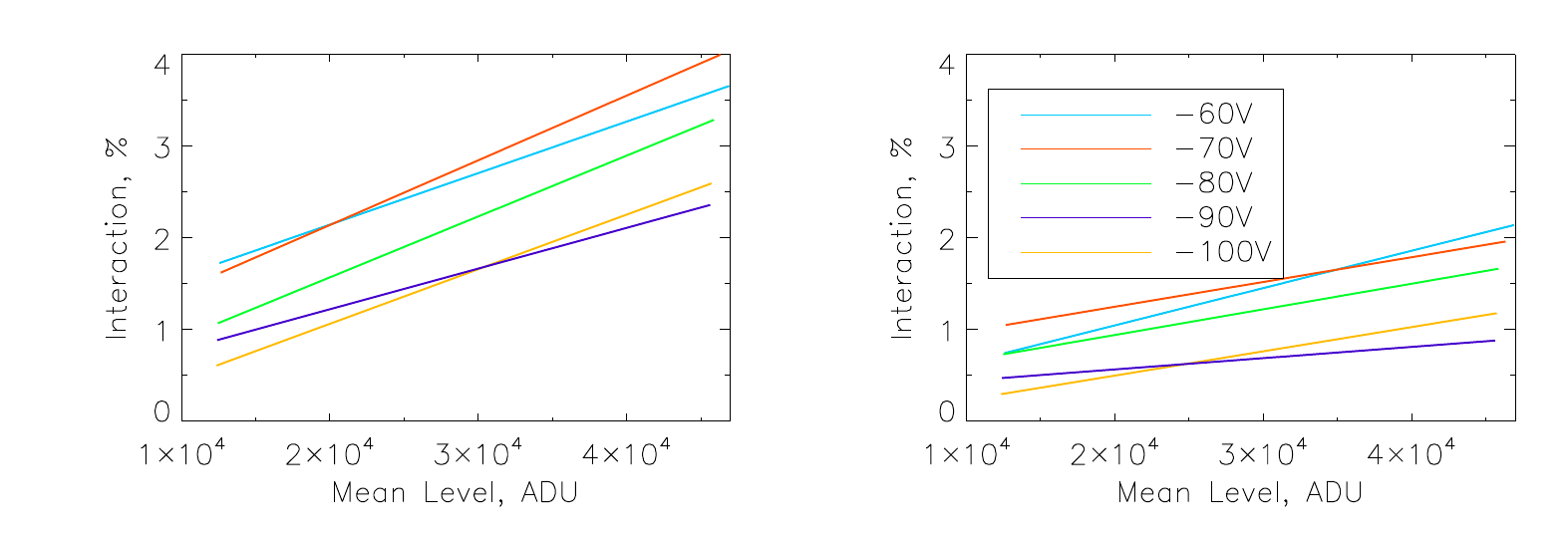}
	\caption{The correlation coefficients $R_{0,1}$, $R_{1,0}$ between pixels in the vertical (left) and horizontal (right) directions versus mean signal level at different voltages of the back substrate of CCD261-84 for wavelength $\lambda=700 nm$}
\end{figure}

Another focus of the work is the development of two high-speed CMOS systems based on $4K\times4K$ GSense4040 and $6K\times6K$ GSense6060 low-noise CMOS detectors.

To work with CMOS sensors a new controller is designed which performs:

\begin{itemize}
\item[--] production of multiple supply voltages;
\item[--] generation of control signals;
\item[--] thermoelectric cooling of the detector;
\item[--] reception and processing of multichannel video data stream;
\item[--] transfering of digital images to the computer.
\end{itemize}


The CMOS system includes a camera, a power supply, and a communication adapter. The design of the camera includes a gas-filled optical head with a CMOS detector and an electronics module combined with an air-liquid cooling system. The housing for electronic modules is made in accordance with the IP66 protection class. Working ambient temperature range is from $-30$ to $+30^\circ C$. The~power supply can be placed at the distance up to 2 meters from the camera. 

The data from the CMOS camera are sent to a computer through the standard 10/40 Gbit Ethernet interface with a fiber-optic communication line at the distance up to 200 m.

The software for data acquisition and processing (\cite{Afanasieva2015}) is programmed using VC++ and QT and operates under  Windows 10 x64. The software provides the following capabilities:

\begin{itemize}
\item[--] control of CMOS system, exposure parameters setup;
\item[--] visualization, analysis and storage of video data;
\item[--] interactive and automatic observation modes;
\item[--] telemetry and diagnostics of the CMOS system;
\item[--] software development kit (SDK).
\end{itemize}

\begin{table}
\caption{CMOS camera system specification for high (HG) and low (LG) gain outputs}
\begin{center}
	\begin{tabular}{l@{\quad}rl}
		\hline
		\multicolumn{1}{l}{\rule{0pt}{12pt}Parameter}&\multicolumn{2}{l}{GSense4040CMT}\\[2pt]
		\hline\rule{0pt}{12pt}
		\hspace{-1mm}Image size, $pixels$  & $4096\times4096$& \\
		Pixel size, $\mu$m  & $9\times9$ & \\
		Quantum efficiency, $\%$  & $74 @600nm$& \\
		Readout noise, $e^-$  & 4.4 (HG), 34 (LG)  & \\
		Nonlinearity, $\%$  & 1.2 (HG), 0.5 (LG) & \\
		Dark current, $Ke^-/pix/s$  & 0.14 (HG), 0.05 (LG) & \\
		Working temperature  & & \\
		\hspace{5mm}-- liquid cooled & $-5 ^\circ C$ & \\
		\hspace{5mm}-- air cooled & $-25 ^\circ C$ & \\
		Interface & 10 Gbit Ethernet & \\
		Camera dimensions, $mm$  & $\diameter135\times240$ &  \\
		Camera weight, $kg$  & 3.5 &  \\
		[2pt]
		\hline
	\end{tabular}
\end{center}
\end{table}

The characteristics of the system with the GSense4040 detector are summarized in Table 2. Geometric noise is measured for this camera system. The amplitude of geometric noise in the bias frame is from 2 to~4~$e^-$. This noise can be corrected by subtracting the averaged bias frame.

The GSense6060-based camera system is currently in the manufacturing process. The backside illuminated scientific sensor provides QE up to 95\%@580nm.

\paragraph{Conclusion.}
The developed CCD systems are expected to be able to increase the sensitivity performance of the BTA spectrographs in the red and near-infrared spectral ranges significantly. Due to the high speed CMOS cameras will be used to investigate rapidly changing processes.

\paragraph{Acknowledgment.}
This work was partly supported by the project of the Ministry of Science and Higher Education of the Russian Federation (including agreement No 05.619.21.0016, project ID RFMEFI61919X0016), and the Fundamental Research Program no. 28 of the Presidium of the RAS (Outer Space: Studies of Fundamental Processes and Their Interrelations). 	

\bibliographystyle{aa}
\bibliography{template}
\end{document}